\documentclass[prodmode,acmtoit]{acmsmall} 

\usepackage[ruled]{algorithm2e}

\SetAlFnt{\small}
\SetAlCapFnt{\small}
\SetAlCapNameFnt{\small}
\SetAlCapHSkip{0pt}
\IncMargin{-\parindent}

\usepackage{graphics}
\usepackage{tikz}

\begin{document}

	\markboth{B. Leiding and A. D{\"a}hn}{Dead Letters to Alice - Reachability of E-Mail Addresses in the PGP Web of Trust}

	\title{Dead Letters to Alice\\Reachability of E-Mail Addresses in the PGP Web of Trust}
	\author{Benjamin Leiding, benjamin.leiding@stud.uni-goettingen.de
	\affil{University of G{\"o}ttingen}
	Andreas D{\"a}hn, andreas.daehn2@uni-rostock.de
	\affil{University of Rostock}}
	
	\date{2016}

		
	\begin{abstract}
		Over the last 25 years four million e-mail addresses have accumulated in the PGP web of trust. In a study each of them was tested for vitality with the result of 40\% being unreachable. Of the mailboxes proven to be reachable, 46.77\% turn out to be operated by one of three organizations. In this article, the authors share their results and challenges during the study.
	\end{abstract}
	
	\maketitle
	
		
	\section{Introduction}
		PGP is a system for (e-mail) encryption and/or signing which is based on asymmetric cryptography. Every PGP user needs at least one key pair (i.e. a public and a private key), which can be used then with software compatible to RFC 4880 \cite{rfc4880} (in the following, ``PGP'' will be used as a synonym for all compatible software, e.g. GPG, PGP, ...) to encrypt, decrypt, sign or verify signed data. Since one of PGP's main use cases is its application to e-mails, key pairs are typically tied to their owners' e-mail addresses. Starting with the use of public key servers to aid users in exchanging public key information \cite{horowitz1997}, e-mail addresses started to accumulate. However, there is no need for publishing a key to a public keyring; public keys are uploaded either on purpose or by accident. Moreover, there is no causality between technical reachability of an e-mail address assigned to a key and the key's usage for encryption or signing purposes; keys without valid or reachable e-mail address can be used without limitation regarding arbitrary data (including e-mails). Users may create a number of public/private key pairs featuring the same e-mail address but each key pair can only be associated with exactly one e-mail address. The data entered during the creation of a key pair (names, e-mail address, ...) are not verified in any way -- neither on creation nor when a public key is added to a key server's database. There is no technical need for this e-mail address to be correct or (publicly) reachable since the PGP web of trust is about exchanging public keys, not e-mail addresses.

		Users sign other users' keys to certify authenticity, thereby building the PGP web of trust. Signing another person's key is broadly understood as a confirmation that the name and e-mail address embedded in the public key match the owner's name and e-mail address. By interpreting this ``signed-by''-relation as a directed edge and each key as a node, the PGP web of trust can be represented as a directed graph. This understanding has originally been introduced by Phil Zimmerman with the manual of PGP 2 \cite{Zimmerman1992}.

		Due to time and space constraints, users of the PGP web of trust cannot verify all the keys they want to use for communication by themselves. Thus, the use of trust chains through parts of the PGP web of trust has become a crucial part of the system. Nevertheless, trustworthiness of a given public key and its binding to an e-mail address is hard to estimate -- as detailed in the following section \ref{s:motivation}.
	
		In October 2014, the PGP web of trust contained roughly four million e-mail addresses which were tested for technical reachability.
	
		Although numerous studies \cite{drewStreibArchiveWebsite:Online}\cite{pgpWOTAnalysis1996:Online}\cite{pgpWOTAnalysis:Online}\cite{JCwotsapp:Online}\cite{hppAnalysisWOT:Online}\cite{MbauerWOTstats:Online} have researched the process of establishing trust relations between PGP web of trust members, little effort has been put into research targeting the currentness of data found in the network.
	
		While the basic idea of testing currentness of e-mail addresses sounds straight-forward, its implementation turned out to entail some complications which are described in this article as well as the study's findings.

		
	\section{Trust metrics}
		\label{s:motivation}
		In the past, several trust metrics were proposed and deployed on the underlying graph of the PGP web of trust, yet they all haven't considered whether e-mail addresses involved in a specific trust chain are still reachable or dead (from a purely technical perspective). Examples for trust metrics developed in the past include: Finding trusted (shortest) paths between two keys \cite{ashley1999gnu}, the mean-shortest-distance \cite{drewStreibArchiveWebsiteMSD:Online}\cite{pgpWOTAnalysis:Online} or betweenness centrality \cite{MbauerWOTstats:Online}\cite{MbauerCentrality:Online}. In addition, the largest connected component of the PGP web of trust, called strong set, is often analyzed in terms of which key is contained in it and its size \cite{hppAnalysisWOT:Online}. All mentioned approaches above exclude expired or revoked keys and signatures, but so far none of these approaches considered excluding keys bound to unreachable email accounts. The implications of revoked, expired keys or signatures and keys (or signatures) bound to unreachable accounts are the same, since they all should be considered as dead and therefore not be used in any kind of trust metric calculation. The underlying reason for such consideration is straight-forward: How shall trust be expressed regarding a communication peer which is inept to communicate? In case that a large percentage of email accounts is not reachable anymore, the usefulness of the approaches mentioned above has to be reconsidered.		
		This paper does not aim for such reconsideration of existing trust metrics as it focuses on the identification of dead e-mail addresses in the PGP web of trust. However, there are some estimations how the findings may influence metrics to be found in the very last section.
	
	
	\section{Study design: Approach and limitations} 
		The study design can be divided into five parts: Preparation, syntax, DNS, validation, and testing. 
	
		As a first step, the keyring data was extracted by means of replicating an existing keyring server. Since an e-mail address may be assigned to multiple keys, duplicates were removed from the sample. The next step included filtering for syntactical validity and removal of obviously bogus addresses. In step 3, the e-mail addresses host portions were tested by pulling each domain's DNS record to extract the corresponding mail exchange (MX) server which is technically responsible for receiving mails addressed to its domain. As an intermediate step, each of these MX servers was tested for validation support. Finally, those e-mail addresses associated with MX servers offering validation were tested one by one.
	
		False negative results may be caused by side effects of the study duration (data becoming outdated). In such cases, data which was cached during early stages of the study was no longer accurate in later test stages, resulting e.g. in failures when trying to connect to (outdated) mail servers.
	
		False positive results might be caused by changes of the e-mail server's validation policy: If an e-mail server was assumed to be validating given e-mail addresses and later (between test stages) its configuration was updated to treat all e-mail addresses as valid, those addresses would be wrongly counted as existing.
	
		Such artifacts might also be caused by e-mail servers which misconceived the study's tests as malicious behavior and therefore tried to mitigate it. If an IDS (intrusion detection system) realized the study testing for existence of e-mail accounts (which is probably something spam bots might do as well), it might have changed the e-mail server's behavior either to non-validation or (more devious since hardly noticeable) random behavior in case of non-existing mailboxes.
	
		A different kind of false positive result is caused by accounts which were technically existing but no longer used. An e-mail address might not be used any longer but still be 	existent from a technical perspective. Without access to the e-mail server itself, this kind of dead mailbox cannot be distinguished from a mailbox which is in actual usage. Thus, in the study's context, flagging an e-mail address as alive does not imply that it's actively used. It's just a sign of the mailbox being technically reachable. Fortunately, unreachable mailboxes left no margin for error: They were unreachable and cannot be used for communication under any circumstances.
	
		In order to assess the influence of the time required for the study's execution on the quality of the results, a follow-up test was performed. It is detailed in Section \ref{secQuality}.
	
		
	\section{Technical Details}
	An e-mail address may be dead for three major reasons: It may be syntactically incorrect, its domain may be invalid or unreachable, or it does not exist on the host.
	
		
		\subsection{Unique and syntactical correct addresses}
			As a first step, double and syntactically incorrect addresses were removed from the test sample. These addresses were excluded from further evaluation since they obviously could never have been used for e-mail communication. Reoccurring e-mail addresses were also removed from the list of addresses to be tested further. 
		
			
		\subsection{Checking the domain}
			In spite of being syntactically correct, an e-mail address may be unreachable due to its domain which might be unaccessible from the public Internet. Domains can be unavailable for numerous reasons: They can be bogus (as in ``foo@bar''), they can be unreachable from the public Internet (private domains or overlay network domains as in ``foo@bar.home'' or ``foo@bar.onion'') or the domain may not be registered any longer. In this study, all domains used for e-mail addresses had their DNS record pulled and tested. E-mail addresses displaying one of the issues described were considered dead as they were not reachable without further effort. Although it seems amiss not to test ``.onion''-addresses (since the effort of connecting with the TOR overlay network is arguably low), it was an essential dogmatic decision of the study design to treat them as unreachable: If .onion addresses were to be considered with special care, every other private network would have needed the same attention.
			E-mail addresses whose domain record does not indicate a valid mail exchange server were marked as done after this step and removed from the list of addresses for further testing. 
		
		
		\subsection{Checking the server behavior}
			Servers may decide to accept e-mails addressed to any recipient in order to guard information about the existence of individual user accounts or to provide a universal fallback address for e-mails with typographic errors in the mailbox portion. While this decision may be worth a try in fighting Spam or lost e-mails, it is contradictory to the study aim. In order to raise efficiency of study realization, each mail exchange host was tested for its address validation policy previous to the actual testing of e-mail addresses. An example for a server offering validation and a server denying validation is provided in Figure \ref{figSMTPexample}: After a connection to the target SMTP server was established, the delivery of an e-mail was attempted. The sender address was set to ``wottest@'' and the actual host name of the machine on which the test was conducted. The receiver address was constructed according to the domain handled by the tested SMTP server; the user account was 27d89e25a3518f4a7434474c2a7d4f1e43911bc58bec5f1 (which was assumed to be non-existent on the target system). If the server accepts the delivery attempt, it was assumed to use a non-validation policy. If the e-mail was rejected, the server was assumed to validate incoming e-mail addresses. There is of course the probability of a false positive, as an account named 27d89e25a3518f4a7434474c2a7d4f1e43911bc58bec5f1 might actually have existed on the tested servers. 
	
	        \begin{figure}[hbt]
	        	\begin{center}
	            	\begin{footnotesize}
	                	\begin{verbatim}
							[ad001@catalina ~]$ host -t mx fbi.gov
							fbi.gov mail is handled by 10 smtpc.fbi.gov.
							[ad001@catalina ~]$ nc smtpc.fbi.gov 25
							220 smtpc.fbi.gov ESMTP
							helo catalina.informatik.uni-rostock.de
							250 smtpc.fbi.gov
							mail from: wottest@catalina.informatik.uni-rostock.de
							250 sender <wottest@catalina.informatik.uni-rostock.de> ok
							rcpt to: 27d89e25a3518f4a7434474c2a7d4f1e43911bc58bec5f1@fbi.gov
							550 #5.1.0 Address rejected.
							quit
							221 smtpc.fbi.gov
							[ad001@catalina ~]$ host -t mx cia.gov
							cia.gov mail is handled by 10 mail2.cia.gov.
							cia.gov mail is handled by 20 mail1.cia.gov.
							[ad001@catalina ~]$ nc mail2.cia.gov 25
							220 mail2a.cia.gov ESMTP
							helo catalina.informatik.uni-rostock.de
							250 mail2a.cia.gov
							mail from: wottest@catalina.informatik.uni-rostock.de
							250 sender <wottest@catalina.informatik.uni-rostock.de> ok
							rcpt to: 27d89e25a3518f4a7434474c2a7d4f1e43911bc58bec5f1@cia.gov
							250 recipient <27d89e25a3518f4a7434474c2a7d4f1e43911bc58bec5f1@cia.gov> ok
							quit
							221 mail2a.cia.gov
							[ad001@catalina ~]$
	                    \end{verbatim}
					\end{footnotesize}
				\end{center}
	            \caption{Example for a SMTP server which validated the local part of the recipient mailbox (fbi.gov) and a SMTP server which did not indicate a non-existent local part of the recipient mailbox (cia.gov); assuming the non-existence of a mailbox named 27d89e25a3518f4a7434474c2a7d4f1e43911bc58bec5f1@cia.gov. Note that no mail was sent to recipients either way as the submission was canceled before any e-mail content was added.}
	            \label{figSMTPexample}
	        \end{figure}
	
			The technical realization of the test ensured that no mail was generated, i.e. no unnecessary mails were sent at any time.
	
			After this step, e-mail addresses belonging to servers which did not allow remote-validation of accounts were marked as inconclusive and removed from the list of e-mail addresses which were subject to further investigation.
	
			
		\subsection{Testing individual e-mail addresses}
			Following the DNS and the server behavior tests, each of the remaining e-mail addresses was subject to an individual test for its existence; two main strategies were used to optimize the process: Reuse of existing TCP connections and parallel processing of queries to different mail exchange servers.
	
			As SMTP relies on TCP, each test would consume not only the overhead of establishing a TCP connection with its three-way handshake, but also raise the risk of triggering an IDS. 
			To avoid such detection and to accelerate the testing process, multiple e-mail addresses were tested using the same TCP connection as tests can be sequenced by issuing the ``RSET'- command to reset the state of the mail exchange server. Further acceleration resulted from simultaneous connections to different servers. Some servers however implement the ``RSET''-command identical to ``QUIT'' and close the connection; they had to be handled separately
	
			Several e-mail servers use gray listing to mitigate junk mail, thus in the study the behavior of a legitimate sender server had to be imitated. Re-transmission of the testing e-mail was tried four times, spanning over a duration of six hours.
	

		
	\section{Data base and Findings}
		Of the 3,183,950 unique e-mail addresses to be found in the keyring snapshot as of 2014-10-22, 2,202,579 were subject to individual validation in which 1,104,029 mailboxes were found to be reachable. This number represents 34.6\% of the overall unique e-mail addresses to be found in the data sample while it represents 50.1\% of the e-mail addresses which were available for testing after all. 892,142 mailboxes (of those available for testing) turned out to be non-existing (representing 28.0\% of all unique e-mail addresses in the sample, respectively 40.5\% of the testable addresses). At the time the snapshot was taken, the PGP web of trust contained 4,388,587 e-mail addresses in total of which 3,183,950 were unique and syntactical correct addresses. As syntactical invalid addresses by definition cannot be used for e-mail exchange, they were not further considered in the study.
	
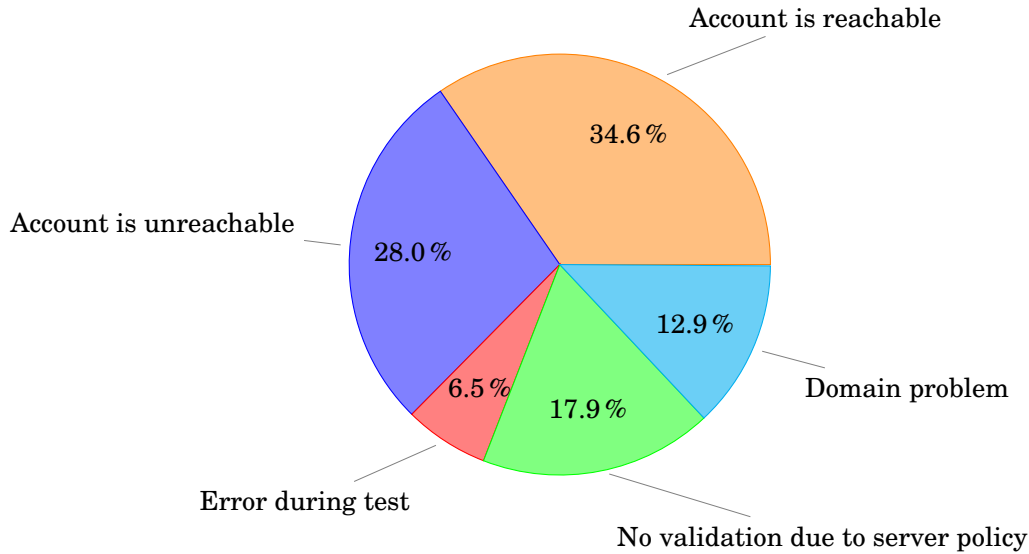
\begin {figure}
			\caption{Pie chart representation of the overall findings for the complete PGP web of trust}
			\label{figPieChart}
			\centering
			
			\def\angle{0}
			\def\radius{2.8}
			\def\cyclelist{{"orange","blue","red","green","cyan"}}
			\newcount\cyclecount \cyclecount=-1
			\newcount\ind \ind=-1
			\begin{tikzpicture}[nodes = {font=\rmfamily}]
			  \foreach \percent/\name in {
			      34.6/Account is reachable,
			      28.0/Account is unreachable,
			      6.5/Error during test,
			      17.9/No validation due to server policy,
			      12.9/Domain problem			      
			    } {
			      \ifx\percent\empty\else               
			        \global\advance\cyclecount by 1     
			        \global\advance\ind by 1            
			        \ifnum4<\cyclecount                 
			          \global\cyclecount=0              
			          \global\ind=0                     
			        \fi
			        \pgfmathparse{\cyclelist[\the\ind]} 
			        \edef\color{\pgfmathresult}         
			        \draw[fill={\color!50},draw={\color}] (0,0) -- (\angle:\radius)
			          arc (\angle:\angle+\percent*3.6:\radius) -- cycle;
			        \node at (\angle+0.5*\percent*3.6:0.7*\radius) {\percent\,\%};
			        \node[pin=\angle+0.5*\percent*3.6:\name]
			          at (\angle+0.5*\percent*3.6:\radius) {};
			        \pgfmathparse{\angle+\percent*3.6}  
			        \xdef\angle{\pgfmathresult}         
			      \fi
			    };
			\end{tikzpicture}
	
		\end{figure}

		The non-existing accounts were joined by all those accounts which are not available for other reasons, mainly DNS (which makes another 12.9\% of the overall unique e-mail addresses). Figure \ref{figPieChart} provides an overview of the study results for the set of syntactical correct e-mail addresses found in the PGP web of trust; detailed results can be found in Table \ref{tblSummary}.
	
		\begin{table}[hbt]
			\tbl{Result summary for the complete PGP web of trust\label{tblSummary}}{
				\begin{tabular}{ll||r|r|r}
					~															&~				& Absolute	&	Relative	&	Relative\\\hline
					\multicolumn{2}{l||}{Syntactical correct unique e-mail addresses in sample}	& 3,183,950	&	100.0\%		&	\\
					\multicolumn{2}{l||}{Filtered out by DNS criteria}							&   410,363	&	 12.9\%		&	\\
						& Domain ceased to exist												&   231,875	&	  7.3\%		&	\\
						& Domain had no MX information											&   160,815	&	  5.0\%		&	\\
						& No DNS information available (DNS timeout$^\ast$)						&    17,651	&	  0.6\%		&	\\
						& Invalid host name length												&        22	&	  0.0\%		&	\\
					\multicolumn{2}{l||}{Domain exists and had MX entry}							& 2,773,587	&	 87.1\%		&	\\
					\multicolumn{2}{l||}{Server allowed no validation$^\ast$$^\ast$}				&   571,008	&	 17.9\%		&	\\
					\multicolumn{2}{l||}{Server allowed validation}								& 2,202,579 &	 69.1\%		&	100.0\%\\
						& Existing mailboxes													& 1,104,029 &	 34.6\%		&	 50.1\%\\
						& Non-existing mailboxes												&   892,142	&	 28.0\%		&	 40.5\%\\
						& Error$^\ast$$^\ast$$^\ast$											&   206,408	&	  6.5\%		&	  9.4\%\\
				\end{tabular}
			}
			\begin{tabnote}%
				\centering
				\Note{$^\ast$}{timeout in recursive queries}
				\vskip2pt
				\Note{$^\ast$$^\ast$}{including unreachable servers}
				\vskip2pt
				\Note{$^\ast$$^\ast$$^\ast$}{technical error during test run, mostly connection timeouts}
			\end{tabnote}%
			
		\end{table}

		There are 571,008 e-mail addresses (resembling 17.9\% of the syntactic correct unique addresses in the sample) which are not accessible for validation. Yet there is only few reason not to assume the deviation in this set to diverge from the results  gathered for e-mail addresses which can be tested. Thus, the amount of alive e-mail addresses can be assumed to be roughly 1.4 millions, making up 44.8\% of the syntactical correct unique e-mail addresses in the data sample.
	
		The testing procedure started with DNS testing on 2015-02-12 and lasted until finishing individual e-mail addresses on 2015-07-24.

		
	\section{Data Quality Sub-study}
		\label{secQuality}
		Because the actual testing took place over a longer period of time, the data quality needed to be evaluated. There are several time-based technical artifacts which may have influenced the results, most notably DNS changes resulting in MX information being outdated or server configuration changes. In order to evaluate these influences, a sub-study was conducted.
	
		To assess the quality of the major study's findings, 1\% of the syntactical correct unique e-mail addresses (31,840) were randomly selected in order to re-do all stages of the study. Afterwards, these results were compared to the major study's results to provide a data quality estimation.
	
		The sub-study shows that 87.7\% of the re-tested e-mail addresses yield the same result as in the overall test, while 12.3\% of the tested e-mail accounts changed their status. While only 161 addresses changed their status from verifiable and existing to verifiable and dead, 465 addresses could not be validated any longer since their mail exchange servers wouldn't allow for such validation, probably due to their operators' fight against unsolicited Spam mails and undesired account verification. 564 addresses which were accessible to be tested for existence were no longer testable due to technical problems. Notably, 141 addresses recognized as dead could be verified as existing in the data quality sub-study. 
		The rest of the changes was due to technical artifacts such as TCP connection errors turning into DNS resolution errors, or misconfigured graylisting (announced temporary unavailability turning out to be rather permanent) -- or vice versa.   
	
		The sub-study took place between 2015-08-11 and 2015-08-30.


	\section{Lessons learned during the realization of the study}
		During the realization of the study, the authors stumbled across several pitfalls of which the four most notable are described in the following subsections. 
	
			
		\subsection{Network access}
			A key necessity to validity of the study's outcome is the ability of a test routine to connect reliably to arbitrary hosts, i.e. a neutral network. Although this topic seemed to be a non-issue in the beginning, it became relevant in sight of whose e-mail addresses may be present in the PGP web of trust: Spammers, scammers, and criminals. Thus, institutional IDS may block, delay or modify data exchanged with servers on blacklists. In fact, it was impossible to check the list of MX servers for blocked entries or guarantee an unlimited network access using the DFN (Deutsches Forschungsnetz; German research network, a network commonly operated by German Universities). This demand for network neutrality may at first glance seem slightly excessive and far-fetched, but actual incidents experienced by the authors make neutrality violations based on connection targets or exchanged contents a realistic concern in study design even in academic environments. Thus, tests were executed using a host within a data center located in Frankfurt/Main (Germany) without institutional IDS, at least for the very first well-known routing hops.

			
		\subsection{Resource problems}
			A naive implementation of the testing approach would consume an unreasonable amount of time. Testing one e-mail address can be assumed to take an average of 15 seconds (not all e-mail addresses are hosted on servers with a network uplink as Google, Microsoft, or Yahoo can afford). A simple approach to test each of the 3.1 million syntactic valid unique addresses thus would take about 553 days. Note that this calculation assumes every TCP connection to be established; waiting for TCP connection timeouts elongates the duration significantly. Consequently, some optimizations are inevitable. Massive straight-forward parallelization however is not an option since it may trigger an IDS. One way to solve this resource problem is the use of the ``RSET''-mechanism as sketched in the technical details.
	
			Distribution of the algorithm over more than one machine seems to be a solution, e.g. using PlanetLab or some cloud service. However, such solutions may suffer from the necessary neutral network access as sketched in the subsection above: In the best case such restrictions are not documented, in the worst case connections to suspicious mail exchange servers are technically restricted and/or considered a violation of the terms of usage.  

			
		\subsection{Server behavior}
			To avoid unnecessary TCP connects (which could trigger an intrusion detection system and which consume extra time), several e-mail addresses belonging to the very same host can be tested without disconnect by issuing the ``RSET''-command. However, some servers implement it identical to a ``QUIT''-command and disconnect. Thus, such optimization cannot be used universally. 
		

			
	\section{Inspecting the PGP web of trust}
		\label{secOverallStats}
		An interesting aspect of the PGP web of trust is the distribution of its users and involved service providers. This section presents statistics based on the host portion of the e-mail addresses, the involved mail exchange servers, and their operating organizations. It combines the number of affected addresses in the web of trust's syntactical correct addresses with the study findings regarding unreachable and reachable addresses.\\
	
\begin{figure}[hbt]
			\centering
			\includegraphics[scale=0.17]{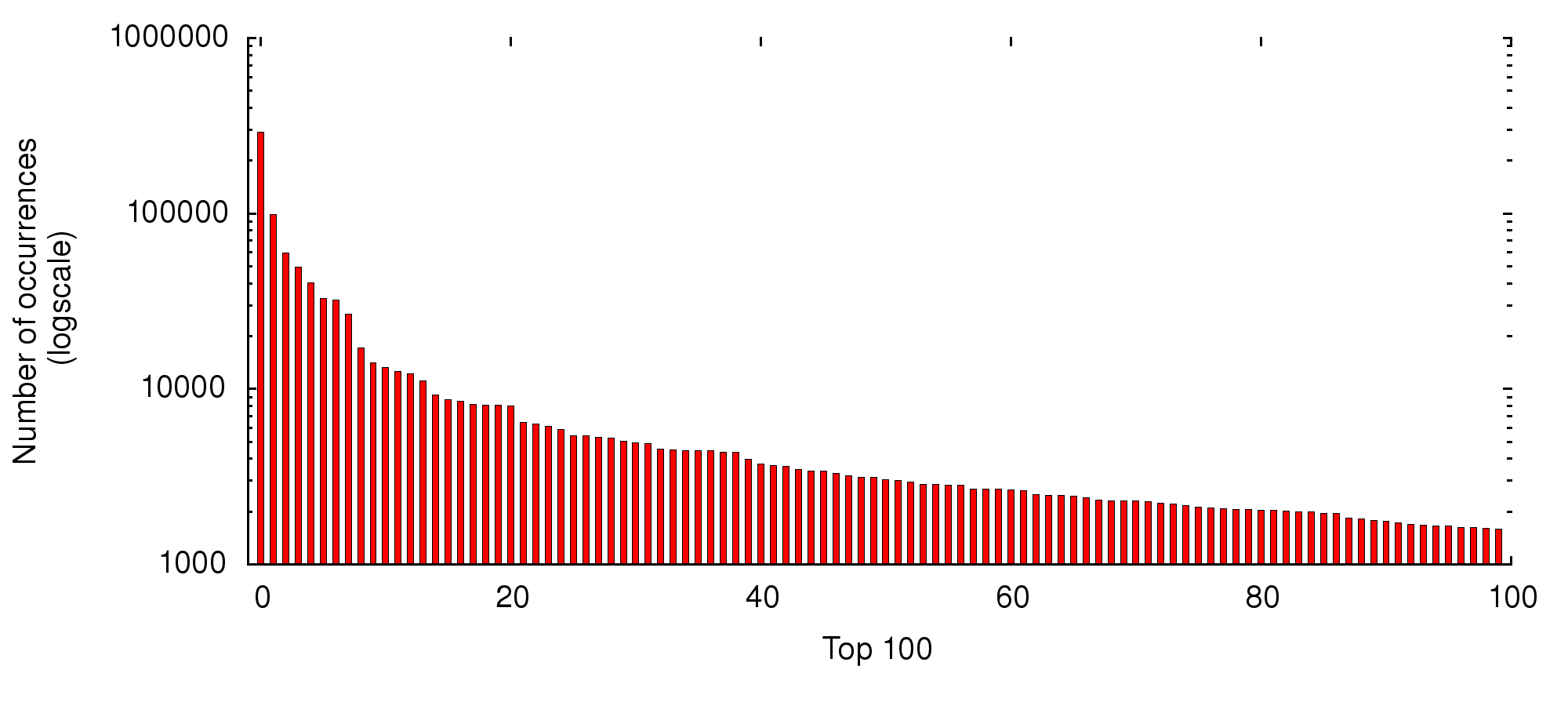}
			\caption{Number of occurrences of hosts present in the PGP web of trust - Top 100}
			\label{fig:distribution_hostbased-top100}
		\end{figure}

		Figure \ref{fig:distribution_hostbased-top100} shows the distribution of users across one hundred e-mail domains with the most users, Table \ref{tblStatHostbased} provides detailed information on the top ten e-mail domains. The number of users per host resembles a power law distribution. The most popular e-mail hosts throughout the PGP web of trust are Google Mail (which is used both with gmail.com and googlemail.com addresses) and Microsoft's e-mail service hotmail.com.
	
		\begin{table}[hbt]
			\tbl{Top 10 hosts present in the PGP web of trust and their study results\label{tblStatHostbased}}{
				\begin{tabular}{l||rr|rr|rr}
					Hostname				&	\multicolumn{2}{|c|}{Overall}	&	\multicolumn{2}{|c|}{Tested dead}	&	\multicolumn{2}{|c}{Tested alive} \\
										&	abs.		& rel$^+$	&	abs.		& rel$^+$$^+$		&	abs		& rel$^+$$^+$$^+$ 	\\ \hline 
					gmail.com				&	290,164		& 9.11 \%		&	11,422		& 1.28 \%		&	278,717		& 25.25 \%		\\	
					hotmail.com				&	98,747		& 3.10 \%		&	39,820		& 4.46 \%		&	40,870		& 3.70 \%		\\
					gmx.de					&	59,281		& 1.83 \%		&	8,625		& 0.97 \%		&	37,391		& 3.39 \%		\\
					yahoo.com$^\ast$			&	49,423		& 1.55 \%		&	n/a		& n/a			&	n/a		& n/a \%		\\
					web.de					&	40,414		& 1.27 \%		&	7,947		& 0.89 \%		&	23,338		& 2.11 \%		\\
					t-online.de				&	32,849		& 1.03 \%		&	4,927		& 0.55 \%		&	9,113		& 0.89 \%		\\
					gmx.net					&	32,136		& 1.01 \%		&	6,989		& 0.78 \%		&	17,876		& 1.62 \%		\\
					aol.com					&	26,814		& 0.84 \%		&	11,272		& 1.26 \%		&	4,136		& 0.37 \%		\\
					googlemail.com				&	17,080		& 0.54 \%		&	805		& 0.09 \%		&	16,273		& 1.47 \%		\\
					home.com$^\ast$$^\ast$			&	14,066		& 0.44 \%		&	14		& 0.02 \%		&	n/a		& n/a \%		\\
				\end{tabular}
			}
			\begin{tabnote}%
				\centering
				\Note{$^+$}{to number of syntactic correct e-mail addresses in PGP web of trust}
				\vskip2pt
				\Note{$^+$$^+$}{to number of e-mail addresses found unreachable in the study}
				\vskip2pt
				\Note{$^+$$^+$$^+$}{to number of e-mail addresses found reachable in the study}
				\vskip2pt
				\Note{$\ast$}{no validation of mailboxes available}
				\vskip2pt
				\Note{$^\ast$$^\ast$}{14,052 validation attempts resulting in connection errors.}								
			\end{tabnote}%
		
		\end{table}
	
\begin{figure}[hbt]
			\centering
			\includegraphics[scale=0.17]{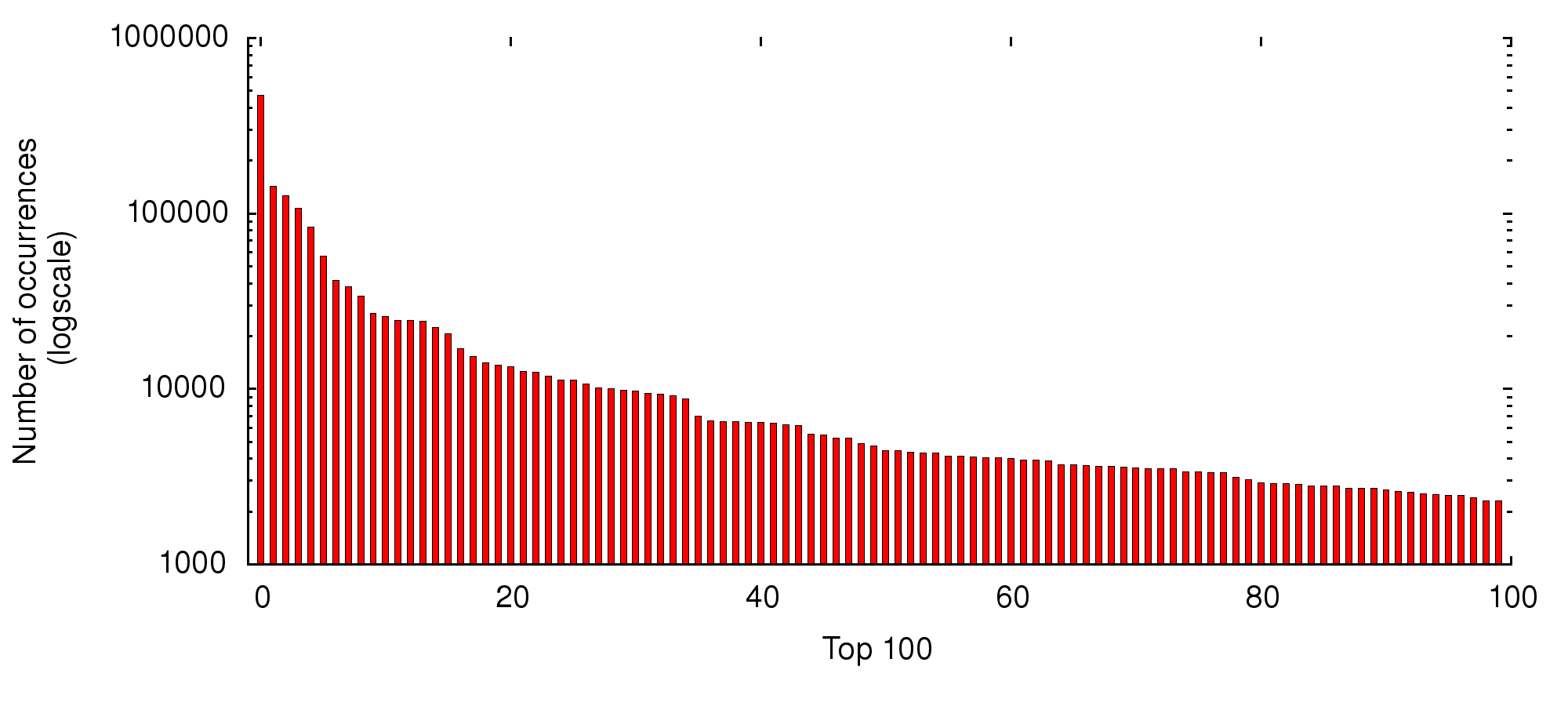}
			\caption{Number of occurrences of mail exchange domains - Top 100}
			\label{fig:distribution_mxbased-top100}
		\end{figure}

		As e-mail exchange servers can also be used from different domains, the host portion of e-mail addresses does not reveal the whole picture. To gain further insight, the distribution of mail exchange servers as found in the DNS records has been investigated. Figure \ref{fig:distribution_mxbased-top100} displays the distribution of users over the topmost one hundred e-mail exchange servers' domains (mapping both mail1.cia.gov and mail2.cia.gov to cia.gov). Details regarding the top 10 MX server domains are provided by Table \ref{tblStatMXbased}. A typical constellation leading to this effect is the use of e.g. Google Mail services with a custom domain. Thus an e-mail address john@doe.gov may lead to an e-mail sent to john's account at the doe.gov domain being technically sent to a Google mail exchange server due to the MX entry of the DNS record.\\ 
	
		\begin{table*}[hbt]
			\tbl{Top 10 mail exchange domains; e-mail addresses which led to errors in DNS resolution or which featured an invalid MX host are left out\label{tblStatMXbased}}{			
				\begin{tabular}{l||rr|rr|rr}
					MX domain 				&	\multicolumn{2}{|c|}{Overall}	&	\multicolumn{2}{|c|}{Tested dead}	&	\multicolumn{2}{|c}{Tested alive}	\\
										&	abs.		& rel$^+$	&	abs.		& rel$^+$$^+$  		&	abs		& rel $^+$$^+$$^+$		\\ \hline
					google.com				&	472,528		& 14.84 \%		&	64,268		& 7.20 \%			&	369,816		& 33.50 \%			\\
					googlemail.com				&	142,350		& 4.47 \%		&	44,401		& 4.98 \%			&	64,599		& 5.85 \%			\\
					hotmail.com				&	125,857		& 3.95 \%		&	50,036		& 5.61 \%			&	53,128		& 4.81 \%			\\
					gmx.net					&	106,818		& 3.35 \%		&	18,943		& 2.12 \%			&	63,796		& 5.78 \%			\\
					yahoodns.net				&	83,747		& 2.63 \%		&	745		& 0.08 \%			&	476		& 0.04 \%			\\	
					outlook.com				&	57,247		& 1.80 \%		&	4,040		& 4.53 \%			&	4,232		& 0.38 \%			\\
					web.de					&	41,402		& 1.30 \%		&	8,262		& 0.93 \%			&	23,834		& 2.16 \%			\\
					aol.com					&	38,393		& 1.21 \%		&	18,897		& 2.12 \%			&	5,003		& 0.45 \%			\\
					t-online.de				&	33,690		& 1.06 \%		&	5,058		& 0.57 \%			&	9,369		& 0.85 \%			\\
					icloud.com				&	26,870		& 0.84 \%		&	2,826		& 0.32 \%			&	24,044		& 2.18 \%			\\
					
				\end{tabular}
			 }
			\begin{tabnote}%
				\centering
				\Note{$^+$}{to number of syntactic correct e-mail addresses in PGP web of trust}
				\vskip2pt
				\Note{$^+$$^+$}{to number of e-mail addresses found unreachable in the study}
				\vskip2pt
				\Note{$^+$$^+$$^+$}{to number of e-mail addresses found reachable in the study}							
			\end{tabnote}%
		\end{table*}
		
		Finally, the question whose servers are involved when mails between accounts present in the PGP web of trust are exchanged shall be considered, as those mail exchange server operators may have a good chance to gain more insight about the PGP web of trust users in terms of meta data analysis. Within the scope of this article, this question cannot be answered in depth since the necessary data is buried in economic constructions such as parent and child companies. However, some companies including brands owned by them or their subsidiaries as known to the authors are outlined in Table \ref{tblStatCompBased}. Based on restricted knowledge available to the authors regarding companies providing e-mail services, the resulting numbers show that at least 46\% of all reachable accounts present in the PGP web of trust are in fact operated (at least in terms of e-mail acceptance) by one out of three organizations: Google, United Internet, or Microsoft.\\ 
	
		\begin{table*}[hbt]
			\tbl{Selected top mx operating companies, including all their brands known to the authors\label{tblStatCompBased}}{		
				\begin{tabular}{l||rr|rr|rr}
					Provider				&	\multicolumn{2}{|c|}{Overall}	&	\multicolumn{2}{|c|}{Tested dead}	&	\multicolumn{2}{|c}{Tested alive}	\\
										&	abs.		& rel$^+$	&	abs.		& rel$^+$$^+$ 		&	abs		& rel$^+$$^+$$^+$		\\ \hline
					Google$^\ast$			& 	614,878		& 19.31 \%		&	108,669		& 12.18 \%			&	343,415		&  31.11\%			\\	
					United Internet$^\ast$$^\ast$	&	216,999		& 6.82 \%		&	49,569		& 5.56 \%			&	115,479		& 10.46 \%			\\
					Microsoft$^\ast$$^\ast$$^\ast$&	183,104		& 5.75 \%		&	54,076		& 6.06 \%			&	57,360		& 5.20 \%			\\	
				\end{tabular}
			}
			\begin{tabnote}%
				\centering
				\Note{$^+$}{to number of syntactic correct e-mail addresses in PGP web of trust}
				\vskip2pt
				\Note{$^+$$^+$}{to number of e-mail addresses found unreachable in the study}
				\vskip2pt
				\Note{$^+$$^+$$^+$}{to number of e-mail addresses found reachable in the study}			
				\Note{$^\ast$}{with MX hosts on domains google.com or googlemail.com}
				\vskip2pt
				\Note{$^\ast$$^\ast$}{with MX hosts 1and1.com, web.de, gmx.\{net, com, de\}, kundenserver.de (managed or unmanaged hosting), or schlund.de}
				\vskip2pt
				\Note{$^\ast$$^\ast$$^\ast$}{with MX hosts on domains outlook.com or hotmail.com}				
			\end{tabnote}%
		
		\end{table*}
	
		
	\section{Consequences}
		There are many works on identity manipulation and validation in context with the PGP web of trust. Unfortunately, in those works the reachability of e-mail addresses associated with public keys is typically not considered. Based on the results of the presented study (40\% of the e-mail addresses assigned to the public keys in the web of trust being unreachable), it seems a promising strategy to consider communication reachability as a new criterion. The common trust metrics used to estimate public keys trustworthiness are all based on transitive trust chains (shortest paths between keys, mean-shortest-distance, betweenness centrality, PGP strong set). Considering public keys associated with unreachable e-mail addresses as dead (similar as for revoked and expired keys) and removing such keys will consecutively destroy trust chains and therefore requires a complete review of the trust estimations derived from these metrics.
	
		Further consequences should be drawn from the numbers laid out in Section \ref{secOverallStats}: As a large number of e-mail accounts involved in the PGP web of trust are operated by few companies. The existence of a global observer who could analyze meta data to gain knowledge about usage of specific keys within the PGP web of trust should be considered. Especially, if users save unencrypted copies of their exchanged mails on storage provided by these operators. 
	
	%
	


		
	\flushleft
	
	\bibliographystyle{ACM-Reference-Format-Journals}
	\bibliography{bibliography}


\begin{thebibliography}{00}


\ifx \showCODEN    \undefined \def \showCODEN     #1{\unskip}     \fi
\ifx \showDOI      \undefined \def \showDOI       #1{{\tt DOI:}\penalty0{#1}\ }
  \fi
\ifx \showISBNx    \undefined \def \showISBNx     #1{\unskip}     \fi
\ifx \showISBNxiii \undefined \def \showISBNxiii  #1{\unskip}     \fi
\ifx \showISSN     \undefined \def \showISSN      #1{\unskip}     \fi
\ifx \showLCCN     \undefined \def \showLCCN      #1{\unskip}     \fi
\ifx \shownote     \undefined \def \shownote      #1{#1}          \fi
\ifx \showarticletitle \undefined \def \showarticletitle #1{#1}   \fi
\ifx \showURL      \undefined \def \showURL       #1{#1}          \fi

\bibitem[\protect\citeauthoryear{Ashley, Copeland, Grahn, and Wheeler}{Ashley
  et~al\mbox{.}}{1999}]%
        {ashley1999gnu}
{J Ashley}, {Matthew Copeland}, {Joergen Grahn}, {and} {David~A Wheeler}. 1999.
\newblock {\em {The GNU Privacy Handbook}}.
\newblock The Free Software Foundation.
\newblock


\bibitem[\protect\citeauthoryear{Bauer}{Bauer}{2004}]%
        {MbauerCentrality:Online}
{M. Bauer}. 2004.
\newblock {C}omputing {B}etweenness {C}entrality in the {W}eb of {T}rust.
\newblock \url{http://shoestringfoundation.org/cgi-bin/blosxom.cgi/2004/12/09}.
    (2004).
\newblock
\newblock
\shownote{2014 ({A}ccessed December 12, 2015).}


\bibitem[\protect\citeauthoryear{{B}auer}{{B}auer}{2006}]%
        {MbauerWOTstats:Online}
{M. {B}auer}. 2006.
\newblock {Betweenness Centrality Statistics for the Strong Component}.
\newblock \url{http://pestilenz.org/~bauerm/wotstats.html}.   (2006).
\newblock
\newblock
\shownote{({A}ccessed September 26, 2015).}


\bibitem[\protect\citeauthoryear{Callas, Donnerhacke, Finney, Shaw, and
  Thayer}{Callas et~al\mbox{.}}{2007}]%
        {rfc4880}
{J. Callas}, {L. Donnerhacke}, {H. Finney}, {D. Shaw}, {and} {R. Thayer}. 2007.
\newblock {OpenPGP Message Format - RFC 4880}.
\newblock http://www.ietf.org/rfc/rfc4880.txt.   (November 2007).
\newblock
\newblock
\shownote{Updated by RFC 5581.}


\bibitem[\protect\citeauthoryear{{C}ederloef}{{C}ederloef}{2013}]%
        {JCwotsapp:Online}
{J. {C}ederloef}. 2013.
\newblock {Web of Trust Statistics and Pathfinder}.
\newblock \url{http://www.lysator.liu.se/~jc/wotsap/index.html}.   (2013).
\newblock
\newblock
\shownote{({A}ccessed September 26, 2015).}


\bibitem[\protect\citeauthoryear{Horowitz}{Horowitz}{1997}]%
        {horowitz1997}
{Mark Horowitz}. 1997.
\newblock {A PGP Public Key Server}.
\newblock   (1997).
\newblock
\newblock
\shownote{{Master Thesis, MIT}.}


\bibitem[\protect\citeauthoryear{McBurnett}{McBurnett}{1996}]%
        {pgpWOTAnalysis1996:Online}
{N. McBurnett}. 1996.
\newblock {{PGP} {W}eb of {T}rust {S}tatistics of 1996}.
\newblock \url{http://bcn.boulder.co.us/~neal/pgpstat/19961206}.   (1996).
\newblock
\newblock
\shownote{({A}ccessed September 26, 2015).}


\bibitem[\protect\citeauthoryear{{M}cBurnett}{{M}cBurnett}{2011}]%
        {pgpWOTAnalysis:Online}
{N. {M}cBurnett}. 2011.
\newblock {PGP} {W}eb of {T}rust {S}tatistics.
\newblock \url{http://bcn.boulder.co.us/~neal/pgpstat/}.   (2011).
\newblock
\newblock
\shownote{({A}ccessed September 26, 2015).}


\bibitem[\protect\citeauthoryear{{P}enning}{{P}enning}{2015}]%
        {hppAnalysisWOT:Online}
{H.~P. {P}enning}. 2015.
\newblock {Analysis of the Strong Set of the PGP Web of Trust}.
\newblock \url{http://pgp.cs.uu.nl/plot/}.   (2015).
\newblock
\newblock
\shownote{({A}ccessed September 26, 2015).}


\bibitem[\protect\citeauthoryear{{S}treib}{{S}treib}{2002a}]%
        {drewStreibArchiveWebsite:Online}
{{D}. {S}treib}. 2002a.
\newblock {K}eyanalyze.
\newblock
  \url{https://web.archive.org/web/20090703200924/http://dtype.org/keyanalyze/index.php}.
    (2002).
\newblock
\newblock
\shownote{({A}ccessed September 26, 2015).}


\bibitem[\protect\citeauthoryear{{S}treib}{{S}treib}{2002b}]%
        {drewStreibArchiveWebsiteMSD:Online}
{{D}. {S}treib}. 2002b.
\newblock {Keyanalyze - Mean-Shortest-Distance}.
\newblock
  \url{https://web.archive.org/web/20090203235946/http://dtype.org/keyanalyze/explanation.php}.
    (2002).
\newblock
\newblock
\shownote{({A}ccessed November 10, 2015).}


\bibitem[\protect\citeauthoryear{Zimmerman}{Zimmerman}{1994}]%
        {Zimmerman1992}
{Phil Zimmerman}. 1994.
\newblock {PGP 2.X Manual}.
\newblock \url{ftp://ftp.pgpi.org/pub/pgp/2.x/doc/pgpdoc1.txt}.   (1994).
\newblock
\newblock
\shownote{({A}ccessed September 17, 2015).}


\end{thebibliography}

\end{document}